\documentclass[11pt,twoside]{article}
\usepackage{asp2010}
\resetcounters

\begin{document}

\title{Looking for the Wind in the Dust}
\author{S. C. Gallagher,$^1$ J. E. Everett,$^2$ S. K. Keating,$^3$ A. R. Hill,$^1$ and R. P. Deo$^1$\affil{$^1$University of Western Ontario, Department of Physics \& Astronomy}
\affil{$^2$University of Wisconsin, Department of Physics}
\affil{$^3$University of Toronto, Department of Astronomy}}

\begin{abstract} 
The blue-shifted broad emission lines and/or broad absorption lines
seen in many luminous quasars are striking evidence for a broad line
region in which radiation driving plays an important role.  We
consider the case for a similar role for radiation driving beyond the
dust sublimation radius by focussing on the infrared regime where the
relationship between luminosity and the prominence of the 3--5~$\mu$m
bump may be key.  To investigate this further, we apply the 3D
hydrodynamic wind model of \citet{Everett2005} to predict the infrared
spectral energy distributions of quasars.  The presence of the
3--5~$\mu$m bump and strong, broad silicate features can be
reproduced with this dynamical wind model when radiation driving on dust is
taken into account.
\end{abstract}

\section{Introduction}
The torus was introduced by \citet{AntonucciMiller} to explain the
revelation from spectropolarimetry that Type~2 (narrow-line) Seyfert
galaxies can look like Type~1 (broad-line) Seyferts along different
lines of sight.  Since that discovery, the torus' primary purpose has
been to obscure the broad-line region, but it also serves to explain
the near- and mid-infrared (IR) emission observed in the vast majority
of quasars whose spectral energy distributions (SEDs) indicate dust
heated to near sublimation temperatures, presumably by the
accretion-disk continuum.

Typically, the torus has been modeled as a static, axisymmetric
structure \citep[e.g.,][]{PierKrolik} that may have condensations
(clumps or clouds) within it \citep[e.g.,][]{Nenkova2008a}.  A notable
exception to this paradigm is the dusty wind model of
\citet{KoniglKartje}, who started with a magneto-hydrodynamic (MHD)
wind, and added in radiation driving on grains beyond the dust
sublimation radius.  They were motivated in part by the large
scale-height of the obscurer required by the observed Seyfert~1 to
Seyfert~2 ratio, the inevitability of radiation driving in the
presence of such a strong source of UV photons, and the apparent 
luminosity-dependent covering fraction of the torus
\citep{Lawrence1991}.  Since then others
\citep[e.g.,][]{ElitzurShlosman} have also incorporated some wind
component into their models, and/or the effects of IR
radiation-pressure in inflating the dusty medium
\citep[e.g.,][]{Dorod2011a}.  The questions remains: how to test for
the existence of this dusty wind?

In the UV, wind signatures are clearly evident in the broad resonance
lines of high ionization species such as C~{\sc iv}.  While most
apparent in classic P-Cygni-like broad absorption lines
\citep[e.g.,][]{PHL5200_disc}, the asymmetries in the same broad
emission lines are also evidence for winds
\citep[e.g.,][]{Leighly,Kruczek}.  The most natural mechanism for
accounting for the high velocities ($\sim10^4$~km~s$^{-1}$) of the
outflows is resonance-line driving by UV continuum photons.

\begin{figure}[t!]
	\centering
	\includegraphics[scale=0.6]{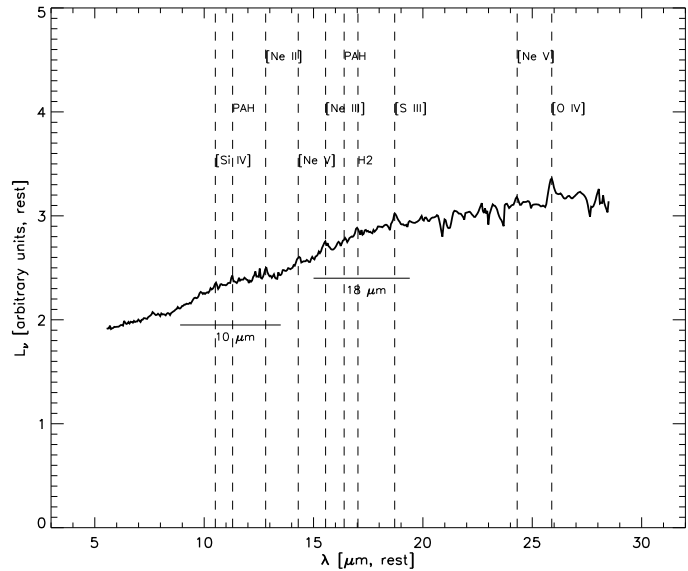}
	\caption{A composite IR spectrum made from $\sim50$ luminous SDSS quasars from the {\em Spitzer} IRS archive.  Prominent emission features are labeled; the horizontal bars indicate the approximate location of the broad silicate emission bumps.  From Hill et al. (in preparation).}
	\label{fig:gall_one}
\end{figure}

Unlike the clear wind signatures in the UV, quasar IR spectra are
characterized by broad features, such as the silicate emission bumps
from the vibrational stretching and bending modes of the grains at 10
and 18~$\mu$m, respectively (see Figure~\ref{fig:gall_one}).  The widths and peaks of
the silicate bumps are affected by the grain size distribution, and so
extracting information on the velocity of the emitting grains is
impossible given the uncertainties in the inherent shape of the bumps.
Narrow atomic features in the IR are from forbidden lines such as
[O~{\sc iv}] and [Ne~{\sc v}] generated in low density,
photoionized gas in the inner parts of the narrow-line region.
Clear spectral signatures of a dusty wind are therefore elusive.

\section{Luminosity Dependence of the 3--5~$\mu$m Bump}

In the UV, the maximum possible terminal velocity of outflows seen in
absorption appears to be set by the UV luminosity
\citep{LaorBrandt,Ganguly}.  This makes sense for a radiation-driven
UV outflow; a high UV photon density has the ability to transfer more
momentum to the ionized wind.  While this particular mechanism is not
relevant in the IR, spectral features in this regime whose strength
depends on luminosity are potentially fruitful for looking for
evidence of the wind from the dust.  One such feature is the
3--5~$\mu$m bump, which can be successfully modeled as a blackbody
with a temperature of 1200---1500~K
\citep[e.g.,][]{Barvainis1987,Deo2011}.  These temperatures are
interesting because they are near the grain sublimation temperature of
graphites, but higher than the grain sublimation temperature of
silicates.  \citet{EdelsonMalkan} first noted that this feature
becomes stronger at higher luminosities.  \citet{Gallagher2007}
confirmed this trend with a much larger sample of SDSS quasars with
{\em Spitzer} IRAC + MIPs photometry by searching for convex spectra
between 1 and 8~$\mu$m; the amount of spectral curvature was significantly
correlated with luminosity.

The increasing prominence of the 3--5~$\mu$m bump can be qualitatively
understood in a wind paradigm, because more luminous quasars will have
more radial IR outflows.  Assuming an accretion disk is the source of
the optical through X-ray continuum, the dust-driving continuum will
always be interior to (and significantly more compact than) the dusty
wind, and thus act in the radial direction.  In this scenario, the
hottest dust is visible from a larger range of inclination angles in
luminous objects.  In the UV, an ionized wind launched from the
accretion disk at small radii ($\sim10^{16}$~cm) could have a
significant vertical component to its acceleration.  This geometric
effect can naturally account for the larger observed fraction of BAL
quasars (seen when looking through the UV wind) compared to
type 2 quasars (seen when the broad-line region is blocked by thick, dusty
material).

\section{The MHD Dusty Wind Model}
To investigate the interplay of radiation driving and the IR SED, we
have developed a dynamical model of the torus as a dusty wind launched
by MHD forces and by radiation pressure from the accretion disk
continuum.  The dusty wind generated in this manner can cover a large
fraction of the sky as seen from the central black hole. This model
has the benefit of being more self-consistent than static torus models
and including the important physics of motion around the black hole
and radiative acceleration. Ultimately, we wish to use this model to
understand how the physical properties of dusty winds in quasars
correlate with their observable spectral signatures in the IR.

The model and its corresponding code has been expanded from its
original form \citep[described in][]{KoniglKartje} by
\cite{Everett2005}, where a comprehensive account of the model's
components and key equations are described. We advance on
\citet{Everett2005} by adding the continuum opacity of ISM dust grains,
as specified by the ISM dust model \citep{Mathis1977,vanHoof2004} in
\verb$Cloudy$ \citep{Ferland1998}.

The radiation pressure on the dust from the central source is very
strong -- even for cases where L/L$_{\rm Edd} = 0.1$, it is
approximately 10 times the force required to unbind dust from the
gravitational potential \citep{Everett2009}.  Dust grains absorb
radially streaming photons originating from the accretion disk, and
then cool by radiating isotropically. The force, felt by the dust
particles due to conservation of momentum, feeds back on the wind
structure by bending the magnetic wind from a vertical to more radial
structure.  The radiative force works in conjunction with the
magneto-centrifugal forces to accelerate the wind flow and therefore
modifies the structure of the outflow.

After iterating the MHD+radiation pressure calculations to set up the
structure of the wind, we use the Monte Carlo radiative-transfer code
MC3D \citep{Wolf2003} to send rays through the wind and predict the
observed IR SEDs.  The input parameters for the wind such as Eddington
ratio ($L/L_{\rm Edd}$), black hole mass ($M_{\rm BH}$), and
illuminating continuum (the empirical optical-to-X-ray composite of
\citealt{Richards2006}, hereafter R06) are varied to investigate their
effects on the output SED.

\clearpage

\begin{figure}[t!]
	\centering
	\includegraphics[scale=0.6]{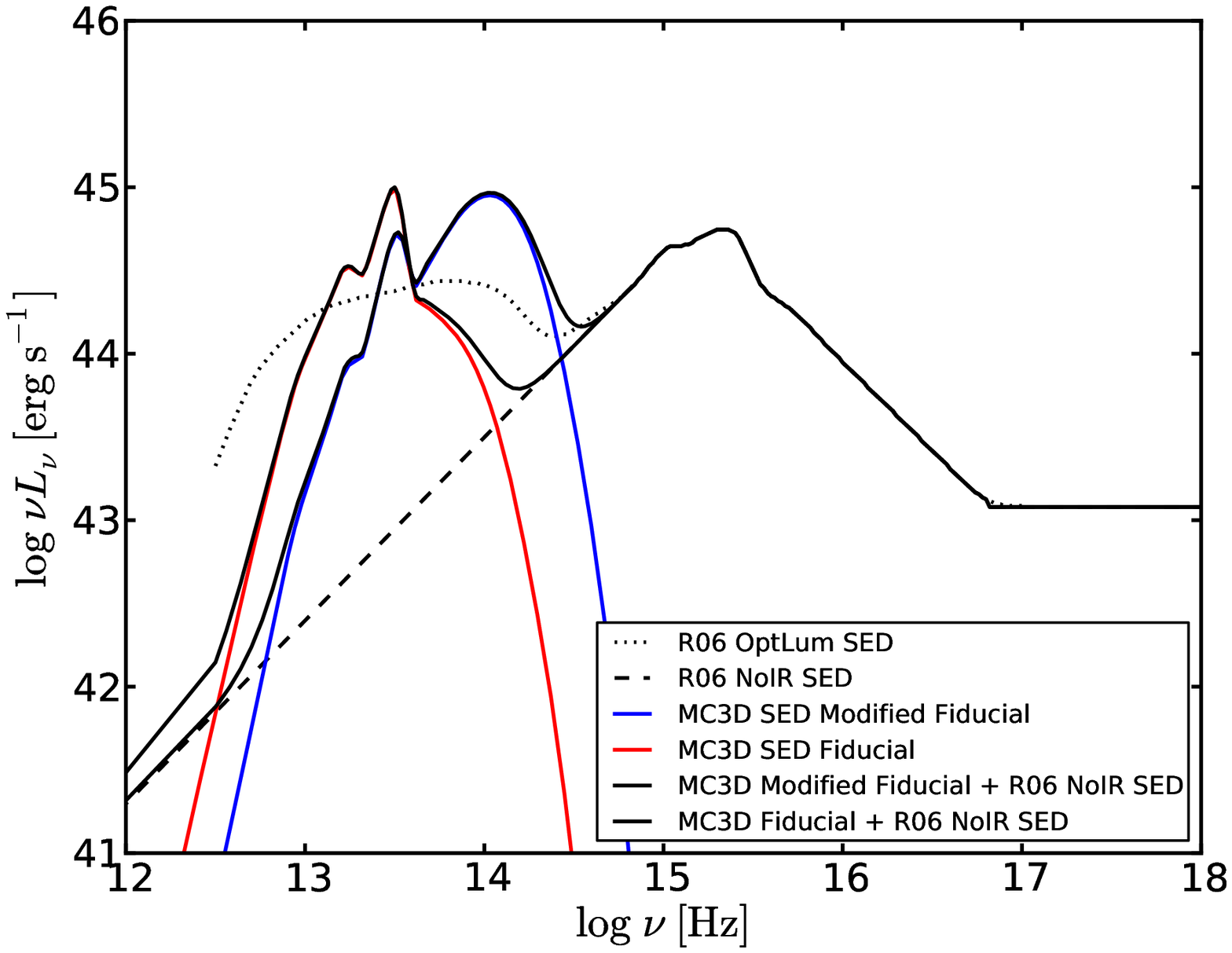}\\
  \includegraphics[scale=0.5]{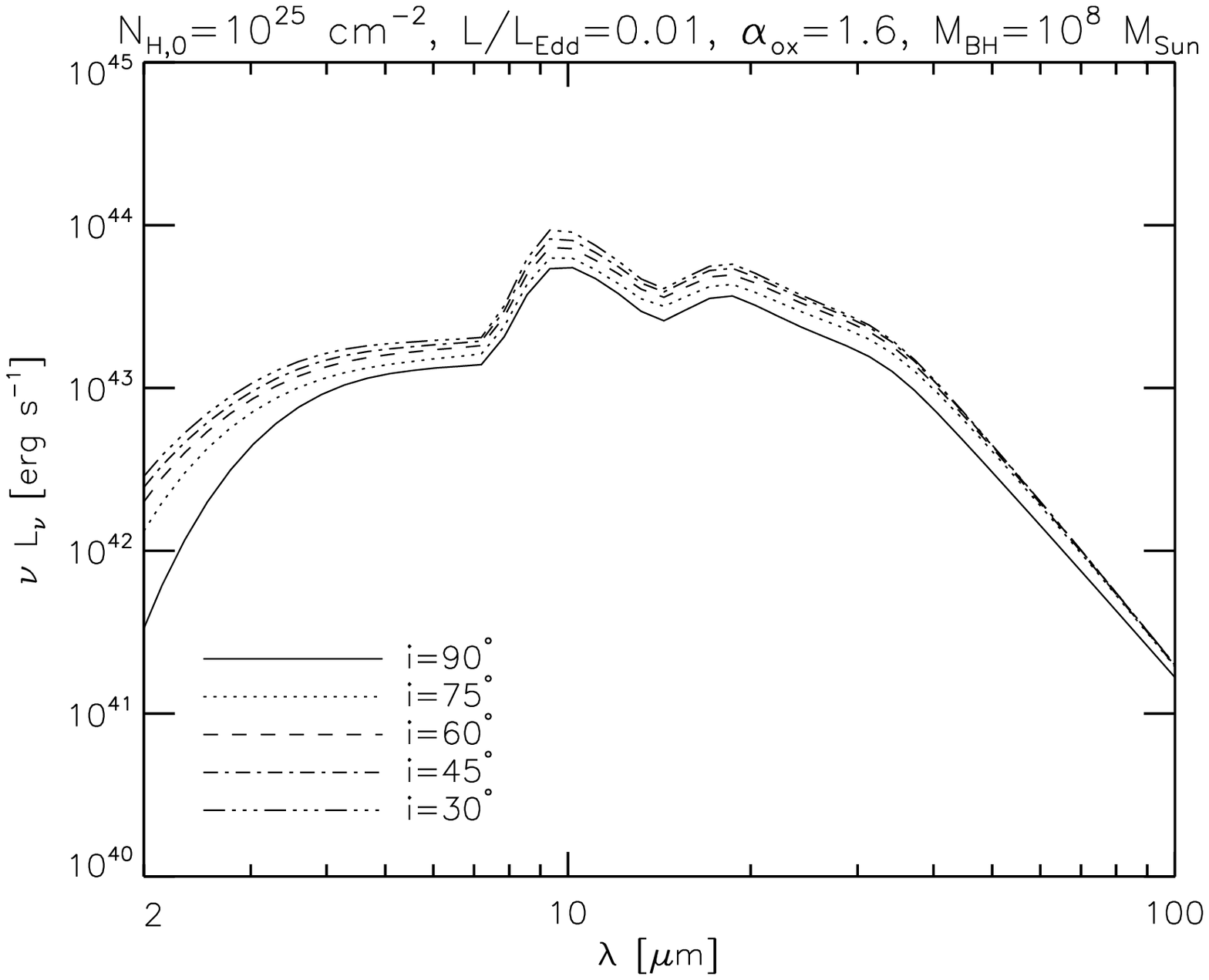}
	\caption{\textsl{Left:} IR to X-ray quasar SEDs as follows: the empirical R06 composite of optically luminous quasars (dotted); the input incident accretion disk continuum (dashed); the output IR SED for the characteristic model (red solid); the summed output IR SED and input incident accretion disk continuum (solid black). A characteristic model with a smaller dust-sublimation radius shows an SED (blue solid) peaking in the near-IR. As the dust-sublimation
    radius is increased the SED power decreases in the near-IR
    and the peak emission shifts to longer wavelengths; small dust
    sublimation radii are required to generate strong 3--5~$\mu$m bump
    emission.  \textsl{Right:} The output IR SEDs generated by the
    wind models; different line styles are used to distinguish the
    observed inclination angle, $i$, with respect to the normal to the
    accretion disk.  The power and shape of the SED is not
    particularly sensitive to the angle, indicating that the wind is
    optically thin to IR radiation for $\lambda\ge 8$~$\mu$m. The
    short wavelength emission increases with smaller inclination
    angle.  This effect could contribute to relatively stronger 3--5 $\mu$m
    emission in more luminous objects because their more radial dusty winds
    reveal the hottest dust in a larger range of inclination angles.
    Figures from Keating et al. (2012).}
	\label{fig:gall_two}
\end{figure}

\clearpage
\section{Principal Results}
With input parameters appropriate for luminous quasars (luminosity of
the incident continuum: $L_{\rm bol}\sim10^{46}$~erg~s$^{-1}$; $L/L_{\rm Edd}
= 0.1$; $M_{\rm BH}$ = 10$^{8}$~M$_\odot$; column density at the base of
the wind: $N_{\rm H,0} = 10^{25}$ cm$^{-2}$; and observer's 
inclination angle from the disk normal: $i = 60^{\circ}$), we can
produce reasonable IR SEDs with approximately the right shape and
luminosity ($L \approx 10^{43} - 10^{45}$ erg s$^{-1}$) as expected
from the R06 composite of optically luminous SDSS quasars
(see~Figure~\ref{fig:gall_two}).  This is a promising result for a
relatively simple model given that we have not attempted any fitting.

A benefit to our model is that we are able to see directly the effects
of various physical parameters on the final IR SED, which ultimately
will allow us to understand the physical properties of the torus
itself. By determining which physical parameters have an observable
effect on the IR SEDs, and narrowing them down so that we generally
reproduce the power expected, we have established a reasonable
starting point from which we can expand and further refine our model.

\section{Future Work} 
In the near future, we plan to implement a more sophisticated
treatment of dust grains that takes into account the different
sublimation temperatures of graphite and silicate.  The luminosity
dependence of emission from the hottest grains --- graphites --- is
promising as a means of demonstrating the connection between radiative
driving of grains, wind geometry, and the observed SED.  We expect
that radiation pressure must be important, because otherwise the
luminosity dependence of the torus covering fraction and of the
3--5~$\mu$m emission is hard to understand.

Our ultimate goal is to generate a library of realistic, IR SEDs that
can be compared with {\em Spitzer} IRS observations of quasars.  We
aim to determine (1) if reasonable input properties (e.g., the input
continuum shape, $L/L_{\rm Edd}$, and $M_{\rm BH}$) result in SEDs
that match in detail those observed, and more ambitiously, (2) how to
use the empirical IR SEDs to independently constrain the unknown
physical properties of quasars.

\acknowledgements{This work was supported by the Natural Sciences and
Engineering Research Council of Canada, the Ontario Early Researcher Award Program, the National Science Foundation, and the {\em Spitzer} Space Telescope Theoretical Research Program.}

%\bibliography{gallagher}

\end{document}